\documentclass[twocolumn,showpacs,preprintnumbers,aps,prb]{revtex4-2}
\usepackage{graphicx}
\usepackage{amsmath,amssymb}
\usepackage{amsthm}
\usepackage{enumerate}
\usepackage{xcolor}
\usepackage{braket}
\usepackage{mathtools}
\usepackage[version=3]{mhchem}
\usepackage[colorlinks=true,linktoc=page,linkcolor=blue,citecolor=red]{hyperref}

\newcommand{\beginsupplement}{%
        \renewcommand{\thetable}{S\arabic{table}}%
        \setcounter{table}{0}
        \renewcommand{\thefigure}{S\arabic{figure}}
        \setcounter{figure}{0}
        \renewcommand{\theequation}{S\arabic{equation}}
        \setcounter{equation}{0}
        \setcounter{section}{0}
        \setcounter{subsection}{0}
     }

\begin{document}
	\title{Schwinger-Boson mean-field study of spin-1/2 $J_1$-$J_2$-$J_{\chi}$ model in honeycomb lattice: thermal Hall signature}
	\author{Rohit Mukherjee, Ritajit Kundu, Avinash Singh and Arijit Kundu}
	\affiliation{Department of Physics, Indian Institute of Technology -
		Kanpur, Kanpur
		208 016, India}

	\begin{abstract}
		We theoretically investigate, within the Schwinger-Boson mean-field theory, the transition from a gapped $Z_{2}$ quantum spin-liquid, in a $J_1$-$J_2$ Heisenberg spin-1/2 system in a honeycomb lattice, to a chiral $Z_{2}$ spin liquid phase under the presence of time-reversal symmetry breaking scalar chiral interaction (with amplitude $J_{\chi}$). We numerically obtain a phase diagram of such $J_1$-$J_2$-$J_{\chi}$ system, where different ground-states are distinguished based on the gap and the nature of excitation spectrum, topological invariant of the excitations, the nature of spin-spin correlation and the symmetries of the mean-field parameters. The chiral $Z_{2}$ state is characterized by non-trivial Chern number of the excitation bands and lack of long-range magnetic order, which leads to large thermal Hall coefficient. 
	\end{abstract}
	
	\maketitle

\section{Introduction}
Quantum spin liquid (QSL) is an exotic state of matter where a spin system does not develop magnetic order nor break any lattice symmetry even at the absolute zero temperature. Instead, the system develops a topological order with fractionalized excitations~\cite{sci,natureb,jins}.  QSLs cannot be described by the traditional Landau's paradigm where different phases are characterized by local order parameters and broken symmetry. Historically QSL was originally proposed by Anderson as a quantum ground state for a geometrically frustrated triangular lattice antiferromagnet~\cite{anderson} and since then, the search for QSL in quantum magnets has primarily focused on the frustrated triangular, Kagome, pyrochlore lattice systems. Among possible candidates, the Kitaev model for spin-1/2 on Honeycomb lattice is a promising candidate to support QSL states, where strong quantum fluctuations arising from bond-dependent interaction destroys the magnetic orders \cite{kita,jins,khau}. This led to experimental search of Kitaev materials and signature of QSL state~\cite{kitexp,kitexo}. In addition to the Kitaev's honeycomb model and the search for Kitaev materials,  the anti-ferromagnetic Heisenberg $J_{1}$-$J_{2}$ model has been studied extensively for possible QSL state~\cite{subir2,subir3}. The conventional ground state of nearest neighbor Heisenberg model (without the $J_2$ coupling), say on the honeycomb lattice, is a N\'{e}el ordered state, but when the second nearest interaction is turned on and increased, the long-range order can get destroyed and the system can enter into a quantum disordered state for intermediate coupling region. Various numerical studies have been conducted that suggests there is QSL phase for intermediate ratio of $J_{2}/J_{1}$ although the parameter range of it has been somewhat debated ~\cite{sondhi1,sondhi2,sondhi3,sondhi4,sondhi5,sondhi7,sondhi16,vmc2}. 
Apart from the novel physics associated with the QSLs, they also hold potentials for applications, especially in the field of quantum information processing~\cite{yang2021probing}, using properties of the long range entangled spins. Such as, the Kitaev QSL can support fractional excitations represented by Majorana fermions~\cite{kitexo}, which can be made to act as anyons obeying non-abelian statistics in the presence of a magnetic field. Braiding these anyons could be an important step toward topological quantum computation~\cite{natureb}. 

\begin{figure}[t]
	\centering
	\includegraphics[width=0.7\linewidth]{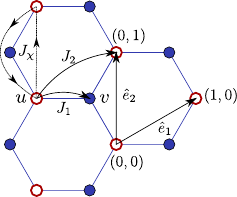}
	\caption{Honeycomb lattice is defined by translation vectors ($\hat{e}_{1}$,$\hat{e}_{2}$) and two sublattices $u$ and $v$. The spin-spin exchange coupling up to second order ($J_1$ and $J_2$, respectively) and chiral coupling, $J_{\chi}$, is depicted.}
	\label{fig:honeycombhopping}
\end{figure}

In recent years, there have also been numerous studies to identify chiral spin liquids (CSL) with realistic spin models having various geometries, such as, the Kagome \cite{kagome1,CSL3,kagome3,kagome4l,kagome5}, triangular \cite{triangle1,triangle2,triangle3,triangle4}, square \cite{square1}, and honeycomb lattices \cite{Haldane}. CSL is a non-magnetic phase, characterized by scaler chiral order (i.e, $\langle \vec{S}_i\cdot(\vec{S}_j\times \vec{S}_k)\rangle \neq0$, where $\vec{S}_i$ is the spin-operator at the $i^{\rm th}$ site) and a finite spin gap. 
%Due to the breaking of TRS the berry curvature is not symmetric under momentum inversion $\Omega_{n\vec{q}} \neq \Omega_{n\vec{-q}}$. This behavior is also carried by the dynamic structure factor $S(\vec{q},\omega) \neq S(\vec{-q},\omega) $. 
The presence of such time-reversal symmetry breaking chiral order can give rise to non-zero chern numbers of the excitations, which can result in enhanced thermal hall conductivity. In particular, a recent work~\cite{kitexo} on the Kitaev honeycomb model predicted that there are two topologically inequivalent phases, within the intermediate disordered regime, one of which is a CSL phase. In that case, the $J_{2}$ interaction in the CSL phase may itself play the same role as a flux term in the Haldane model~\cite{haldane1}, and the $J_{2}$ term acts as a spin-orbital coupling for the spinons in similarity with the Kane-Mele model~\cite{haldane2}. Very Recently authors in the Ref~\onlinecite{Chen} have investigated the topological phase transition and nontrivial thermal Hall signatures in honeycomb lattice magnets in presence of Zeeman coupling using the Abrikosov-fermion mean-field theory, which reports similar findings of unusual thermal Hall effect for pseudogap phase of copper-based superconductors \cite{subir6}.

In the present work, we consider the $J_1$-$J_2$ spin $S=1/2$ Hiesenberg model along with a scalar chiral three-spin term. Without the scalar chiral term, in the classical limit, $S\rightarrow \infty$, the system is N\'{e}el ordered for $J_2/J_1<1/6$ and magnetically ordered in a spiral manner for $J_2/J_1>1/6$~\cite{classical1,classical2,classical3}.  For the  quantum-case, the nature of the ground-state has been extensively studied (without the scalar chiral term), using spin-wave theory~\cite{quantum1,quantum2,classical2,classical3}, non-linear $\sigma$ model~\cite{nonsigma}, exact diagonalization~\cite{ed1,ed2}, variational Monte-Calro~\cite{vmc1,vmc2} and other methods~\cite{other}. The general understanding is that, for $J_2/J_1 \lessapprox 0.21$, it orders magnetically as a N\'{e}el phase; between  $0.37 \lessapprox J_2/J_1 \gtrapprox 0.21$ there is a gapped spin-liquid (GSL) phase; between  $0.4 \lessapprox J_2/J_1 \gtrapprox 0.37$ there is a $C_3$ rotational symmetry broken disordered valence-bond crystal (VBC) state and for $J_2/J_1 \gtrapprox 0.4$, the system orders magnetically in a spiral manner. In the present work, based on the Schwinger-Boson mean-field theory (SBMFT), we investigate the effect of the scalar spin-chiral term on the disordered (gapped) phases and we theoretically observe a transition to a chiral $Z_{2}$ (CZSL) state, where the Chern numbers of the excitation bands change. We also theoretically study its signature in the thermal Hall measurement.

The paper is organized as following. In Sec.~\ref{sec:formalism} we briefly review the formalism of SBMFT and various technicalities involved in solving for the ground-state properties. We provide details of the numerical simulation and further discussion of how we identify various phases from numerical data in Sec.~\ref{sec:numsim} and we present the numerical results in Sec.~\ref{sec:numres}. We discuss the results further and summarize our findings in Sec.~\ref{sec:sum}.

\section{Formalism}\label{sec:formalism}
In this work, we study the effect of the scaler three-spin chiral term, with coefficient $J_{\chi}$, in the $J_1$-$J_2$ Heisenberg spin-1/2 Hamiltonian:
\begin{equation}\label{eq:H}
	H=\sum_{nn}J_{1}\vec{S_{i}} \cdot \vec{S_{j}}+\sum_{nnn}J_{2}
	\vec{S_{i}} \cdot \vec{S_{j}}+ \sum_{\triangle} J_{\chi}
	\vec{S_{i}} \cdot (\vec{S_{j}}\times \vec{S_{k}})
\end{equation}
where $\vec{S}_{i}$ is the spin operator at site $i$, $J_{1}$ and $J_{2}$
are the coupling amplitude for nearest and next nearest neighbors, whereas $J_{\chi}$ is the amplitude of the scalar spin-chiral term, as outlined in the Fig.~\ref{fig:honeycombhopping}. In the third term, the sum involves the triangular plaquettes $\Delta$ formed by the nearest-neighbors, as shown in the same figure. 

As a passing comment, strong coupling expansion of Hubbard model yields $J_{1}=4 t^{2}_{1}/U$ and $J_{2}=4 t^{2}_{2}/U$, where $t_{1}$ and $t_{2}$ are the nearest and next-nearest neighbor hopping amplitudes of electrons, respectively, and $U$ being the onsite repulsion. On the other hand, scalar spin chirality is proportional to $-24t^{2}_{1}t_{2}/U^{2} \ \sin\Phi$, where $\Phi$ is the magnetic flux through the triangular plaquette~\cite{Hubbard}. Starting from the Haldane-Hubbard model, one may also naturally lead to $J_{\chi}$ term without any further application of magnetic field \cite{Haldane,CSL1,Haldane,CSL3,CSL4,Hubbard,CSL6}

We study this spin-model, Eq.~(\ref{eq:H}), using the Schwinger-Boson mean-field theory (SBMFT), where we represent the spin-operators in terms of Bosons. The nature of the bosonic excitations on top of the mean-field ground-state predicts order-disorder transition and other physical properties of the system, as we discuss later. Before we discuss our numerical findings, we present a short review of the SBMFT below.

\subsection{Schwinger-boson mean-field theory}
The principle idea behind SBMFT is to express the spin operators in terms of bosonic operators that carry spin. In the SU(2) representation where two bosonic flavors are introduced to describe the spin operators, we write~\cite{SCQM}
\begin{equation}
	\vec{S_{i}}=\dfrac{1}{2} b_{i,\sigma}^{\dagger}\vec{\tau}_{\sigma\sigma'}b_{i,
		\sigma'}
\end{equation}
where $\tau^i$ are the Pauli matrices, and $ b_{i,\sigma}^{\dagger}$ are the
bosonic creation operator of spin $\sigma$ on site $i$. In order to
preserve the SU(2) commutation rule , the following local constraint has
to be fulfilled on every site:
\begin{equation}
	\label{eq:mu}
	\sum_{\sigma}b_{i\sigma}^{\dagger}b_{i\sigma}=2S.
\end{equation}
Where $S$ is the value of spin under consideration, which we take to be $1/2$ for the present work. However it is typically difficult to impose this constraint exactly~\cite{subir4}; thus we impose it on the average over the mean-field ground-state. 

As we do not impose any symmetry to be broken in the ground-state, only possible bilinears that preserves the spin rotation symmetry are the following:
\begin{equation}
	\hat{A}_{ij}=\dfrac{1}{2}[b_{i\uparrow}b_{j\downarrow}-b_{i\downarrow}b_{j\uparrow}]
\end{equation}
\begin{equation}
	\hat{B}_{ij}=\dfrac{1}{2}[b_{i\uparrow}^{\dagger}b_{j\uparrow}+b_{i\downarrow}^{\dagger}b_{j\downarrow}].
\end{equation}
It is clear that $A_{ij}$s measure singlet type correlations while the $B_{ij}$s measure triplet correlations~\cite{thermal3}. In a gapped phase, the first one is favored, whereas the triplet correlation allows the spinons to hop between sites giving rise to long range orders.

 It can be easily verified that,
\begin{equation}
	\vec{S_{i}} \cdot \vec{S_{j}}=:\hat{B}^{\dagger}_{ij}\hat{B}_{ij}:-\hat{A}^{\dagger}_{ij}\hat{A}_{ij},
\end{equation}
where $:\hat{O}:$ refers to the normal ordering. Now, we perform the mean-field decoupling of $\hat{A}$, $\hat{B}$ operators as
\begin{align}
	&\hat{A}^{\dagger}_{ij} \hat{A}_{ij} \rightarrow
	A^{*}_{ij}\hat{A}_{ij}+\hat{A}^{\dagger}_{ij} A_{ij}-
	A^{*}_{ij}A_{ij},\\
	&\hat{B}^{\dagger}_{ij} \hat{B}_{ij} \rightarrow
	B^{*}_{ij}\hat{B}_{ij}+\hat{B}^{\dagger}_{ij} B_{ij}-
	B^{*}_{ij}B_{ij},\label{eq:mfdecomp}
\end{align}
with $A$, $B$ are the mean-field order parameters that are computed, self-consistently, from the average over the mean-field ground-state, $\ket{\rm gs}$,
\begin{equation}
	\label{eq:ABavg}
	A_{ij}= \braket {{\rm gs}|\hat{A}_{ij}|{\rm gs}}, \ \ B_{ij}= \braket
	{{\rm gs}|\hat{B}_{ij}|{\rm gs}}.
\end{equation}
 These expectation values, collectively define the parameters of the mean-field ansatz.
The expectation values are calculated in the new basis that digonalizes the Hamiltonian and using,
	\begin{equation}\label{eq:gammags}
	\gamma_{\vec{q},\lambda} \ket{\rm gs}=0,
	\end{equation}
where, $\ket{\rm gs}$ is the vacuum state for the resulting bosonic excitation, $\gamma_{\vec{q},\lambda}$, details of these procedure  we shall discuss in the sub-section~\ref{sec:mfd}. Once the decomposition Eq.~(\ref{eq:mfdecomp}) is done, the effective mean-field Hamiltonian is now completely expressed in terms of bosonic bilinears. In the same way we can do the SBMFT decoupling of the scaler chirality term where we use the following identity,
\begin{equation}
	\vec{S_{i}}.(\vec{S_{j}}\times
	\vec{S_{k}})=2i(-\hat{B}^{\dagger}_{ki}\hat{B}^{\dagger}_{jk}\hat{B}^{\dagger}_{ij}+\hat{B}_{ij}\hat{B}_{jk}\hat{B}_{ki}),
\end{equation}
Which we write, using the mean-field decomposition as,
\begin{equation}
	\begin{split}
		\hat{B}_{ij}\hat{B}_{jk}\hat{B}_{ki} \approx
		\hat{B}_{ij}\braket{\hat{B}_{jk}}\braket{\hat{B}_{ki}}+\braket{\hat{B}_{ij}}\hat{B}_{jk}\braket{\hat{B}_{ki}}+
		\\ \braket{\hat{B}_{ij}}\braket{\hat{B}_{jk}}\hat{B}_{ki}- 
		2\braket{\hat{B}_{ij}}\braket{\hat{B}_{jk}}\braket{\hat{B}_{ki}}.
	\end{split}
\end{equation}

\begin{figure}[t]
	\centering
	\includegraphics[width=0.9\linewidth]{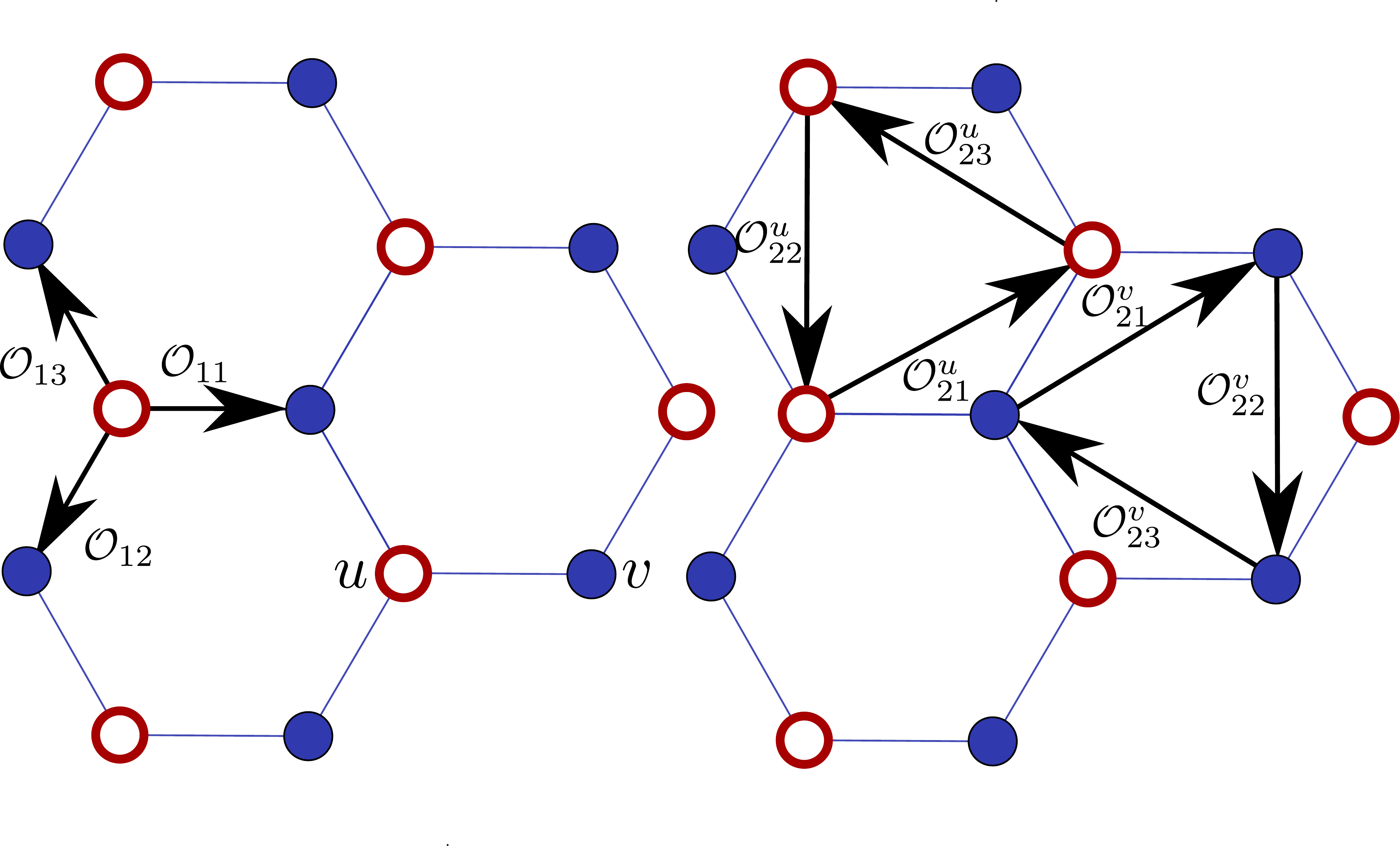}
	\caption{The nine independent mean-field complex parameters
		$\mathcal{O}_{id}$ and their clockwise orientation conventions that alows the point group symmetry breaking. The first subscript $i$ refers to the
		neighbors (nearest, next-nearest), while the second refers to the
		three directions. For the next-nearest neighbors $i = 2$, connected
		sites are on the same sublattices; we then introduce two sets of meanfield
		parameters labeled with the extra superscript $O^{w}$, with $w=u/v$. $\mathcal{O}$ can be of type $A$ or $B$ between the sites, totaling 18 independent parameters.}
	\label{fig:meandieldans}
\end{figure}
In our our original Hamiltonian we have the Heisenberg interactions up to second nearest neighbor. Now, if we want to preserve the translational symmetry but break all the point group symmetries, we can get at most 18 inequivalent mean-field ansatz (bond parameters), 9 for each $A_{ij}$ and $B_{ij}$. These are schematically shown in the Fig.~\ref{fig:meandieldans}. For nearest neighbor interactions, the bonds are between one $u$ to one $v$ sublattices, denoted by $\mathcal{O}_{1d}$ where $d$ are the three possible orientations (subscript 1 denotes nearest neighbor). For the next nearest interactions, the bonds are connections between two $u$ or $v$ sublattices. We denote them with $\mathcal{O}^{w}_{2d}$ where the superscript represents sublattice index and $d$ is the three possible orientations as before. Each $\mathcal{O}$ can be chosen as $A$ or $B$ type of order parameters, totaling 18 of them.

The final mean-field Hamiltonian can be expressed as,
\begin{equation}\label{eq:mfH}
\begin{aligned}
H_{\rm mf}=&\sum_{ij}J_{ij}(-A_{ij}^*\hat{A}_{ij}+B_{ij}^*\hat{B}_{ij}+h.c.)\\
&-\sum_{i}\mu_i\left(\sum_{\sigma}b_{i\sigma}^{\dagger}b_{i\sigma}-2S\right)+2i\sum_{\triangle}J_{\chi}(B_{ki}B_{jk}\hat{B}_{ij}+\\
&B_{ij}B_{ki}\hat{B}_{jk}+B_{ij}B_{jk}\hat{B}_{ki}- h.c)+K
\end{aligned}
\end{equation}\\
with,
\begin{equation}
K=\sum_{ij}J_{ij}(|A_{ij}|^2-|B_{ij}|^2)+8\sum_{\triangle}J_{\chi} \Im(B_{ij}B_{jk}B_{ki}).
\end{equation}
The final term in the Hamiltonian is the consequence of the constraint Eq.~(\ref{eq:mu}). For simplicity, we assume the chemical potential to be either $\mu_u$ or $\mu_v$, depending on the sublattices. Schwarz inequality restricts the upper bounds on the moduli $|A| \leq S+1/2$, $|B| \leq S$, which must be obeyed for any self-consisistent ansatz in SBMFT~\cite{subir4}.

\begin{figure*}[t]
	\centering
	\includegraphics[width=1.0\linewidth]{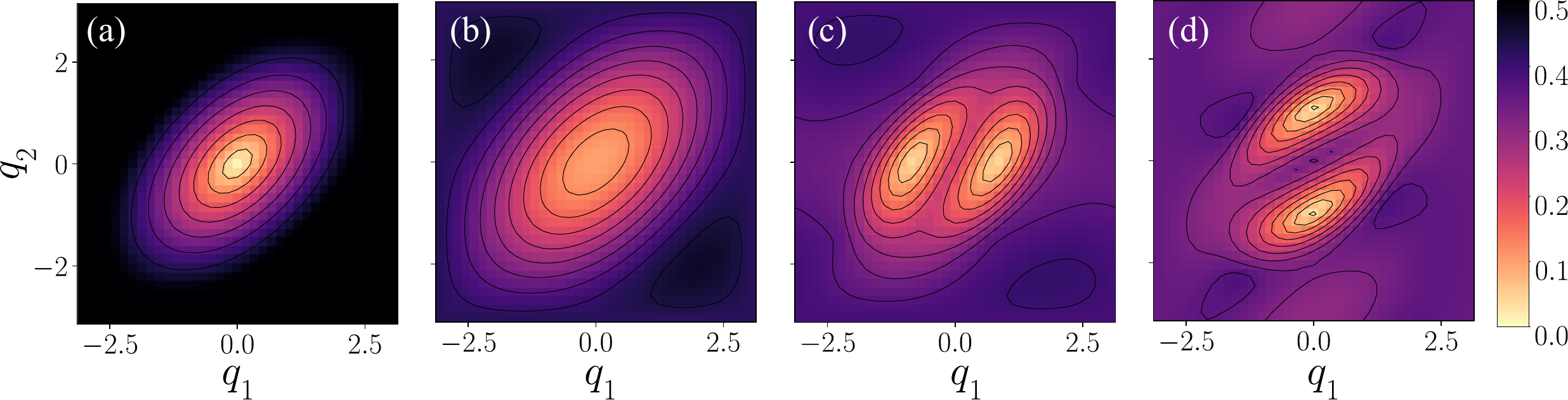}
	\caption{Dispersion of the lowest spinon-band with $J_{\chi}=0$ and $J_1=1$. (a), (b), (c) and (d) are the dispersions of the spin bands, when the system is in the N\'{e}el phase ($J_2=0.197$), GSL phase ($J_2=0.35$), VBC phase ($J_2=0.38$) and spiral anti-ferromagnetic phase ($J_2=0.43$), respectively.}
	\label{fig:dispersion}
\end{figure*}

\subsection{Diagonalization of bosonic quadratic Hamiltonian}

The mean-field Hamiltonian, Eq.~(\ref{eq:mfH}), can be diagonalized using the Bogoliubov-Valantin canonical transformation~\cite{rohit1,diag}. The procedure is following for a generic quadratic bosonic Hamiltonian,
\begin{equation}
	\label{eq:bosonH}
	H=\dfrac{1}{2} \Psi^{\dagger} M  \Psi; \ \Psi^{\dagger}=(b_{1}^{\dagger},...,b_{N}^{\dagger},b_{1},..,b_{N}).
\end{equation}
$N$ is the degree of freedom and $M$ is an $2N\times 2N$ matrix. $b_{n}^{\dagger}$ ($b_{n}$) are the creation (annihilation) operators in  momentum, spin or any other degrees of freedom. In order to find the eigenvectors corresponding to the matrix $M$, we introduce creation (annihilation) operators $\gamma_{m}^{\dagger}$ ($\gamma_{m}$) such that,
\begin{equation}
	\Psi= T \Gamma, \ \ \Gamma^{\dagger}=(\gamma_{1}^{\dagger},...\gamma_{N}^{\dagger},\gamma_{1},...\gamma_{N}),
\end{equation}
where $T$ is the basis-transformation matrix. We choose our $T$ such that Hamiltonian in Eq. (\ref{eq:bosonH}) can be written in a diagonal form as:
\begin{align}
	\begin{aligned}	\label{eq:bosonHD}
	&H=\dfrac{1}{2} \Gamma^{\dagger} T^{\dagger} M T \Gamma,\\
	&{\rm with},~ T^{\dagger}MT=\begin{pmatrix}
		\omega_{1} & 0& \cdots& 0 && \\ 0& \omega_{2}& \cdots&0 &&\\
		\vdots&\vdots& \ddots &\vdots&&\\0 &0 &\cdots& \omega_{2N}
	\end{pmatrix}.
\end{aligned}
\end{align}
In order to preserve the bosonic commutation rules, the $\Psi$ and $\Gamma$ matrices should obey the following matrix equation,
\begin{align}
	[\Psi_{i},\Psi_{j}^{\dagger}]=[\Gamma_{i},\Gamma_{j}^{\dagger}]=(\rho_{3})_{ij},\nonumber
\end{align}
where
\begin{align}
\rho_{3} \equiv \begin{pmatrix}
	I_{N \times N} & 0 && \\ 0 & -I_{N \times N}
\end{pmatrix}.
\end{align}
Here $I_{N \times N}$ is the identity matrix of dimension $N$. This implies that the transformation matrix must satisfy,
\begin{equation}
	T \rho_{3} T^{\dagger}=\rho_{3},
\end{equation}
In a more formal language, $T$ is a paraunitary \cite{diag} SU$(N,N)$ matrix. The elements of the transformation matrix can be found from the eigenvectors of the dynamic matrix, defined as
\begin{equation}
	K=\rho_{3} M
\end{equation} 
which satisfies the Heisneberg equation of motion for $\Psi$~\cite{subir6}. All the eigenvalues of the dynamic matrix (when it is diagonalizable) appears in pairs of opposite sign, and are real. $T$ is also referred as the derivative matrix, consisting of all the eigenvectors of $K$ sorted in the form,
\begin{equation}
	T=[V(\omega_{1}),....,V(\omega_{N}),V(-\omega_{1}),...,V(-\omega_{N})],
\end{equation}
with the eigenvectors normalized as,
\begin{equation}
	V^{\dagger}(\omega_{i})\rho_{3}V(\omega_{i})=1, \ \ V^{\dagger}(-\omega_{i})\rho_{3}V(-\omega_{i})=-1,
\end{equation}
for all the sets of $(V(\omega_{i}),V(-\omega_{i}))$. After the diagonalization we have,
\begin{equation}
	T^{-1}KT= \text{diag}\ (\omega_{1},...,\omega_{N},-\omega_{1},...,-\omega_{N}),
\end{equation}
and
\begin{equation}
	T^{\dagger}MT=\text{diag}\ (\omega_{1},...,\omega_{N},\omega_{1},...,\omega_{N}).
\end{equation}
Both $M$ and $K$ are simultaneously diagonalized. We call the the positive (negative) bands with indices $n=1,...,N$ ($n=N+1,...,2N$) as the particle (hole) bands.\\

\begin{figure}
	\centering
	\includegraphics[width=1.\linewidth]{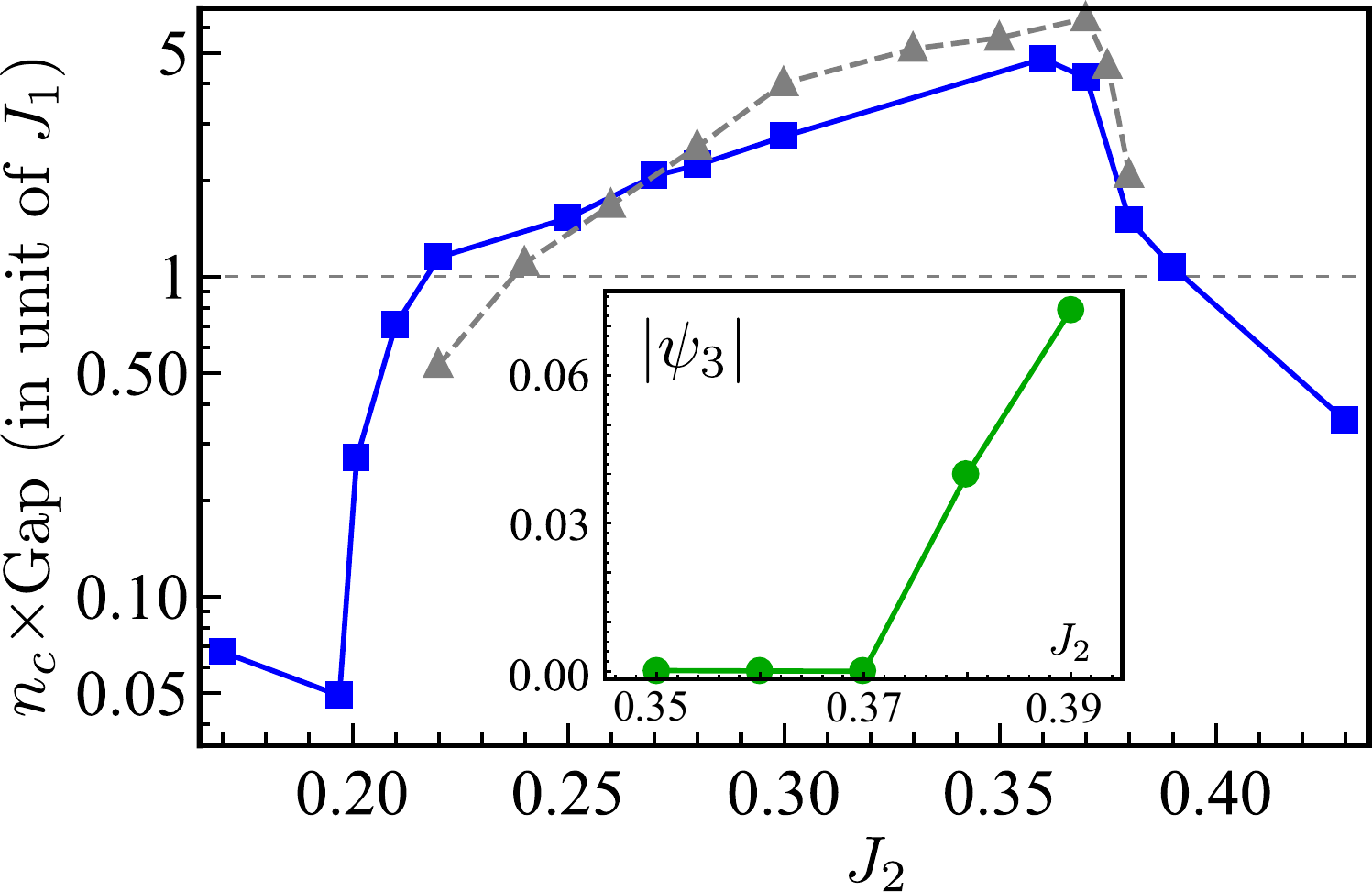}
	\caption{ Blue curve indicate the gap in excitation spectrum as a function of $J_{2}$, with $J_{\chi}=0$, distinguishing gapped and gapless (defined as when the gap is $<1/N$, with $N=36$ in our case) phases. Grey curve corresponds to the same plot with a different system size, $N=48$. Inset: the $C_3$ symmetry breaking order parameter, $\psi_3$, defined in Eq.~(\ref{eq:C3}), becomes non-zero in the small window of $J_2$, when we expect the VBC ground state.}
	\label{fig:gap1}
\end{figure}

\subsection{Mean-field dispersion}\label{sec:mfd}
We use the method of the preceding section for diagonalization of the mean-field Hamiltonian Eq.~(\ref{eq:mfH}), in the momentum space.
% For honeycomb lattice each unit cell contains two sublattice $w =u, v$ and each site is labelled by its unit cell coordinate $r$. 
We write the Bosonic annihilation operator in the Fourier space as,
\begin{equation}
	b_{\vec{r},w,\sigma}= \dfrac{1}{\sqrt{n_{c}}} \sum_{\vec{q}} e^{i
		\vec{q}.\vec{r}} b_{\vec{q},w,\sigma},
\end{equation}
where $n_{c}$ is the total number of unit cells in the real-space lattice (each containing two sub-lattices); $\vec{r}$ are the positions of the unite cells and $w=u,v$  are sub-lattice indices. The combination $(\vec{r},w)$ defines position of a particular site and $\sigma=\uparrow/\downarrow$ are the flavors of the Schwinger-Bosons. Then the
mean-field Hamiltonian is in the momentum space is written as:
\begin{equation}
	H_{\rm mf}=
	\frac12\sum_{\vec{q}}\Psi_{\vec{q}}^{\dagger}M_{\vec{q}}\Psi_{\vec{q}}-(2S+1)n_{c}\sum_{w}
	\mu_{w}+K,
\end{equation}
with,
\begin{equation}
	\Psi_{\vec{q}}^{\dagger}=(b_{\vec{q},u,\uparrow}^{\dagger}
	b_{\vec{q},v,\uparrow}^{\dagger}
	b_{-\vec{q},u,\downarrow}b_{-\vec{q},v,\downarrow} ).
\end{equation}
Where the co-efficient matrix $M_{\vec{q}} = M_{\vec{q}}^{(1)}+M_{\vec{q}}^{(2)}$, consisting of two parts, where the second term is proportional to the scalar chirality $J_{\chi}$. The first of these terms is given by,
\begin{widetext}
\begin{align}
	 M_{\vec{q}}^{(1)}= \begin{pmatrix}
		J_{2}(B_{2d}^{u}\phi_{2d}+B_{2d}^{u*}\phi_{2d}^{*})+2\mu_{u}
		& J_{1} B_{1d}^{*}\phi_{1d}^{*} & J_{2}
		A_{2d}^{u}(\phi_{2d}-\phi_{2d}^{*}) & -J_{1}
		A_{1d}\phi_{1d}^{*} \\
		J_{1} B_{1d}\phi_{1d}& 
		J_{2}(B_{2d}^{v}\phi_{2d}+B_{2d}^{v*}\phi_{2d}^{*})+2\mu_{v}
		& J_{1} A_{1d} \phi_{1d} & J_{2}
		A_{2d}^{v}(\phi_{2d}-\phi_{2d}^{*}) \\
		J_{2} A_{2d}^{u*}(-\phi_{2d}+\phi_{2d}^{*}) & 
		J_{1} A_{1d}^{*} \phi_{1d}^{*} & 
		J_{2}(B_{2d}^{u}\phi_{2d}^{*}+B_{2d}^{u*}\phi_{2d})+2\mu_{u}
		& J_{1} B_{1d}\phi_{1d}^{*}\\
		-J_{1} A_{1d}^{*}\phi_{1d}&
		A_{2d}^{v*}(-\phi_{2d}+\phi_{2d}^{*}) &
		J_{1}B_{1d}^{*}\phi_{1d}  &
		J_{2}(B_{2d}^{v}\phi_{2d}^{*}+B_{2d}^{v*}\phi_{2d})+2\mu_{v}
	\end{pmatrix},\nonumber
\end{align}
\end{widetext}
where we have assumed the summation over repeated index $d$ and $\phi_{id}(\vec{q})= e^{i \vec{q}.\vec{\delta}_{i,d}}$ is the phase factor generated between two neighboring sites at distance $\vec{\delta}_{i,d}$ from $i^{th}$ neighbors (1, 2) and in one of three directions, $d$, shown in Fig.~\ref{fig:meandieldans}.

\begin{figure*}
	\centering
	\includegraphics[width=0.95\linewidth]{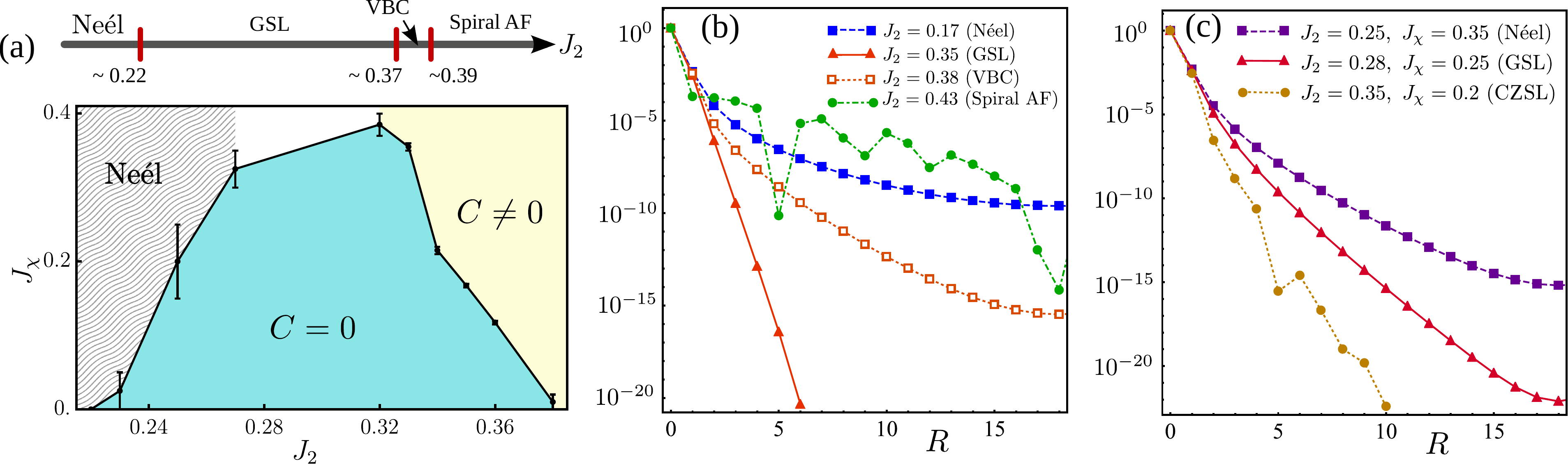}
	\caption{(a) Phase diagram of $J_{1}$-$J_{2}$-$J_{\chi}$ Heisenberg model on the honeycomb lattice. Above: the phase diagram without the spin-chiral term (i.e, $J_{\chi}=0$). Below: the phase diagram with the scalar spin-chiral term, where a gapped chiral spin liquid phase (CZSL) emerges with non-vanishing Chern number of the spinon bands. We set $J_{1}=1$ as our scale of energy. We identify the various phase with the symmetry of the order-parameters (see Sec.~\ref{sec:numsim}), the gap in the spectrum, as well as how the spin-spin correlation function decays. (b) and (c) shows the spin-spin correlation functions $|\langle \vec{\hat{S}}_{0} \cdot \vec{\hat{S}}_{R} \rangle/\langle \vec{\hat{S}}_{0} \cdot \vec{\hat{S}}_{0} \rangle|$, as a function of $R$ (along $\hat{e}_2$) in a logarithmic scale along the vertical axes. (b) shows such correlations without the schiral spin term, $J_{\chi}$, whereas in (c) we show the same with $J_{\chi}\neq0$, for various phases. It is evident that the correlation decays much faster in a GSL, CZSL and VBC phase, whereas the decay is slower for the case of gapless states.}
	\label{fig:phase}
\end{figure*}
The coefficient matrix $M_{\vec{q}}^{(2)}$ is given by
\begin{equation*}
M_{\vec{q}}^{(2)} =	2J_{\chi}\begin{bmatrix}
		-2{\rm Im}C_{11} & C_{12} & 0 & 0 \\
		C_{12}^{*} & -2{\rm Im}C_{22} & 0 & 0\\
		0&0 & -2{\rm Im}C_{33} & C_{34}\\
		0 & 0 & C_{34}^{*} & -2{\rm Im}C_{44}
	\end{bmatrix},
\end{equation*}
$C_{ij}$ are $\vec{q}$ dependent expressions, details of which can be found in appendix.
	\begin{align*}
		&C_{11}=B_{12}^{*} B_{13} e^{-i q_{2}}+ B_{12}^{*}B_{11} e^{-iq_{1}}+ B_{11}^{*} B_{13} e^{i(q_{1}-q_{2})},\\
		&C_{22}=B_{13}^{*}B_{11}e^{i(q_{2}-q_{1})}+B_{11}^{*}B_{12} e^{i q_{1}}+ B_{12}^{*}B_{13}e^{-i q_{2}}.
%		&C_{33}= B_{12}^{*}B_{13}e^{i q_{2}}+ B_{12}^{*} B_{11}e^{i q_{1}}+ B_{11}^{*}B_{13}e^{-i(q_{1}- q_{2})} + {\rm c.c} ,
%		&C_{44}=J_{\chi}[(i B_{13}^{*}B_{11}e^{-i (q_{2}-q_{1})})+(i B_{11}^{*}
%		B_{12} e^{-i q_{1}})+ (i B_{12}^{*}B_{13}e^{i q_{2}} )+ {\rm c.c} ].
	\end{align*}
$C_{33}$ and $C_{44}$ are the same as $C_{11}$ and $C_{22}$, respectively, after the exchange $q_1 \rightarrow -q_1$, $q_2 \rightarrow -q_2$.
	\begin{align*}
&		C_{12}=i(\mathcal{A}-e^{-iq_{1}}\mathcal{B}+e^{-i(q_{1}-q_{2})}\mathcal{C}),\\
&		C_{34}=i(-\mathcal{A}^* + e^{-iq_{1}}\mathcal{B}^*   - e^{-i(q_{1}-q_{2})}\mathcal{C}^*).
	\end{align*}
with,
\begin{align*}
\mathcal{A} &= B_{12}^* B_{21}^{u*}-B_{23}^{u}B_{13}^{*}+B_{23}^{v} B_{13}^{*}- B_{12}^{*} B_{21}^{v*},\\
\mathcal{B} &= B_{22}^{u*}B_{13}^{*}+ B_{21}^{u} B_{11}^{*}  + B_{13}^{*} B_{22}^{v*} -  B_{21}^{v}B_{11}^{*},\\
\mathcal{C} &= B_{12}^{*}B_{22}^{u}+	B_{11}^*B_{23}^{u*}-B_{11}^{*} B_{23}^{v*}+ B_{22}^{v}B_{12}^{*}.
\end{align*}

To find the eigenmodes corresponding to $M$, we introduce new
annihilation (creation) operators $\gamma  (\gamma^{\dagger})$, as before, such that,
\begin{equation}
	\Psi_{\vec{q}}= T_{\vec{q}} \Gamma_{\vec{q}}.
\end{equation}
with,
\begin{equation}
	\Gamma_{\vec{q}}^{\dagger}=(\gamma_{\vec{q},u,\uparrow}^{\dagger}
	\gamma_{\vec{q},v,\uparrow}^{\dagger}
	\gamma_{-\vec{q},u,\downarrow}\gamma_{-\vec{q},v,\downarrow} ).
\end{equation}
Now the mean-field Hamiltonian takes the form, in this new basis:
\begin{equation}
	H_{\rm mf}= \frac12\sum_{\vec{q}}\Gamma_{\vec{q}}^{\dagger} \ \hat{\omega}_{\vec{q}} \
	\Gamma_{\vec{q}}-(2S+1)n_{c}\sum_{w} \mu_{w}+K
\end{equation}
the matrix $T_{\vec{q}}$ satisfies the following conditions,
\begin{equation}
	T_{\vec{q}}^{\dagger} \ \rho_{3} \ T_{\vec{q}}= \rho_{3},
\end{equation}
\begin{equation}
	T_{\vec{q}}^{\dagger} \ M_{\vec{q}} \ T_{\vec{q}}= \hat{\omega}_{\vec{q}},
\end{equation}
where
\begin{equation}
	\hat{\omega}_{q}= I_2 \otimes  \begin{bmatrix} 
		\epsilon_{\vec{q},u} & \\  &\epsilon_{\vec{q},v}
	\end{bmatrix}.
\end{equation}
Now that we have found the mean-field spinon dispersion, one can find the fixed point in the mean-field parameter space by minimizing the free energy,
\begin{equation}\label{eq:dFmf}
	\mathcal{F}_{\rm mf}=\sum_{\vec{q},w}\epsilon_{\vec{q},w}-(2S+1)n_{c}\sum_{w}\mu_{w}+K
\end{equation}
with respect to the mean-field parameters and the chemical potentials:
\begin{equation}
	\dfrac{\partial \mathcal{F}_{MF}}{\partial \mathcal{O}_{id}} =0, \ \ 	\dfrac{\partial \mathcal{F}_{MF}}{\partial \mu_{w}}=0.
\end{equation}
These equations can be solved numerically. In a second procedure, which we employ in the present work, we solve for the mean-field order parameters by self-consistently solving Eq.~(\ref{eq:ABavg}).

\begin{figure}[t]
	\centering
	\includegraphics[width=0.8\linewidth]{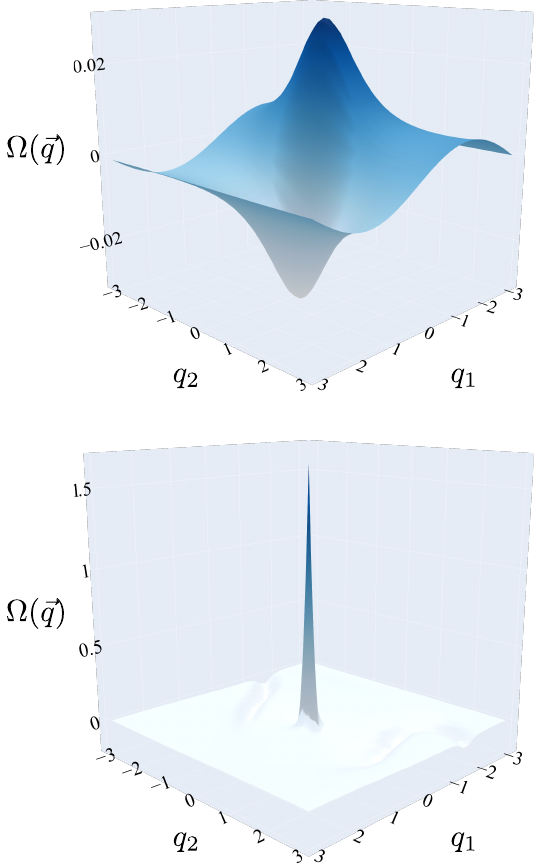}
	\caption{Plot of berry curvature of lower spinon band for $J_{2}=0.35$; with $J_{\chi}=0.15$ (top), which is before the transition to chiral $Z_{2}$ (CZSL), and, $J_{\chi}=0.17$ (bottom), after the transition to CZSL. The value of $\Omega$ is shown in the unit of $1/\delta q_1\delta q_2$, where $\delta q_i$ is the momentum-grid separation along the $i$ direction.}
	\label{fig:berry1}
\end{figure}

\subsection{spin structure factor}\label{sec:ssf}
Although we work in a finite size lattice system, how the static spin-structure factor $ \langle \vec{\hat{S}}_{0}  \cdot \vec{\hat{S}}_{i} \rangle$ behaves as a function of $R_i$ reveals the nature of the underlying ground-state. In such finite size system the spin-rotation symmetry is never broken in the ground-state, which allows us to write
\begin{align}
		& \langle \vec{\hat{S}}_{0}  \cdot \vec{\hat{S}}_{i} \rangle=3\langle \hat{\vec{S}}^{z}_{0} \hat{\vec{S}}^{z}_{i} \rangle \nonumber\\
		&~~~~~~~~=\dfrac{3}{4}\Big \langle \Big( \hat{b}^{\dagger}_{0\uparrow} \hat{b}_{0\uparrow}-\hat{b}^{\dagger}_{0\downarrow} \hat{b}_{0\downarrow}               \Big)\ \Big \langle \Big( \hat{b}^{\dagger}_{i\uparrow} \hat{b}_{i\uparrow}-\hat{b}^{\dagger}_{i\downarrow} \hat{b}_{i\downarrow}               \Big)\rangle.
\end{align}
In the Fourier-space, we have
\begin{align}
		& \hat {S^{z}_{0}}\hat {S^{z}_{i}}=\dfrac{1}{4N^{2}} \sum_{\vec{k},\vec{q},\vec{k'},\vec{q'}} e^{i(\vec{q}-\vec{q'}) \cdot \vec{r}_{i}}\Big[ \hat{b}^{\dagger}_{\vec{k}\uparrow}\hat{b}_{\vec{k'}\uparrow}\hat{b}^{\dagger}_{\vec{q}\uparrow}\hat{b}_{\vec{q'}\uparrow} \nonumber\\
		& +\hat{b}^{\dagger}_{\vec{k}\downarrow}\hat{b}_{\vec{k'}\downarrow}\hat{b}^{\dagger}_{\vec{q}\downarrow}\hat{b}_{\vec{q'}\downarrow}-\hat{b}^{\dagger}_{\vec{k}\uparrow}\hat{b}_{\vec{k'}\uparrow}\hat{b}^{\dagger}_{\vec{q}\downarrow}\hat{b}_{\vec{q'}\downarrow}-\hat{b}^{\dagger}_{\vec{k}\downarrow}\hat{b}_{\vec{k'}\downarrow}\hat{b}^{\dagger}_{\vec{q}\uparrow}\hat{b}_{\vec{q'}\uparrow}\Big],
\end{align}
where we have suppressed the sublattice index for brevity. The expectation values of these operators can be calculated in the diagonal basis of the Hamiltonian and using Eq.~(\ref{eq:gammags})~\cite{ssf3}. 

%The  dynamic structure factor is defined as,
%\begin{equation}
%	S({\vec q},\omega)=\dfrac{1}{N_{s}}\sum_{l,j} e^{i \vec{k} \cdot (\vec{r}_{i}-\vec{r}_{j})} \int_{-\infty}^{\infty} dt e^{-i \omega t}  \langle {\vec\hat{S}}_{l} \cdot {\vec\hat{S}}_{j} \rangle
%\end{equation}
%which using the Bogoluibov operators takes simpler forms from which we calculate the ddf. The static (equal time) structure factor is obtained by integrating over frequencies,

%\begin{equation}
%	S(\vec{q})= \int d\omega S(\vec{k},\omega).
%\end{equation}

\subsection{Berry Curvature and Thermal Hall effect}
Once we diagonalize the bosonic Hamiltonian, we have the Hamiltonian of the excitation
\begin{equation}
	H^{\rm D}=\sum_{\vec{q}}\sum_{n=1}^{N_{\rm band}}\epsilon_{\vec{q},n}\left(\gamma_{\vec{q},n}^{\dagger}\gamma_{\vec{q},n}+\dfrac{1}{2}\right),
\end{equation}
where $N_{\rm band}$ is the number of bosonic particle bands (with $\epsilon_{\vec{q},n}>0$), which is two in our case. The thermal hall co-efficient is then defined as~\cite{shindou2013topological},
\begin{equation}
	\kappa_{xy}=-\dfrac{k_{B}^{2}T}{\hbar
		V}\sum_{\vec{q}}\sum_{n=1}^{N_{\rm band}}\left[c_{2}[n_{B}(\epsilon_{\vec{q},n})]-\dfrac{\pi^{2}}{3}\right]\Omega_{n\vec{q}},
\end{equation}
where $n_{B}(\omega)$ is the Bose distribution function and,
\begin{equation}
	c_{2}(x)=\int_{0}^{x}dt \Big(\ln \dfrac{1+t}{t}\Big)^{2}.
\end{equation}
$\Omega_{n\vec{q}}$ is the Berry curvature in momentum space, for the $n^{\rm th}$ band, defined as.
\begin{equation}
	\Omega_{n\vec{q}}\equiv i\epsilon_{\mu
		\nu} \left[ \rho_{3}\dfrac{\partial\ T_{\vec{q}}^{\dagger}}{\partial
		k_{\mu}}\rho_{3}
	\dfrac{\partial T_{\vec{q}}}{\partial k_{\nu}} \right]_{nn},
\end{equation}
which can also be recasted in the following form,
\begin{equation}
	\Omega_{n\vec{q}}=i \epsilon_{\mu\nu} \bra{\partial_{\mu}\psi_{n}(q)}\rho_{3}\ket{\partial_{\nu}\psi_{n}(q)},
\end{equation}
where $\psi_{n}(k)$ is the $n^{\rm th}$ coloumn of the $T_{q}$ matrix. The numerical evaluation of the Berry curvature follows the U(1)-link variable method, outlined in the Appendix. The Chern number is then evaluated as
\begin{equation}
	C_{n}=\dfrac{1}{2\pi}\int_{BZ}\Omega_{n\vec{q}} \ \text{d}\vec{q},
\end{equation}
which is always an integer and also it obeys the following constraints
\begin{equation}
	\sum_{n=1}^{N_{\rm band}} C_{n}= \sum_{n=N_{\rm band}+1}^{2N}C_{n}=0,
\end{equation}
that is the sum of Chern numbers over particle and hole bands are individually zero \cite{shindou2013topological}.

\section{Details of the numerical simulation}\label{sec:numsim}
We solve for self-consistent values of the mean-field parameters in a finite lattice of $n_c = N\times N$ unit-cells, where $N=36$ (containing $2\times36\times36$ sites). There is numerical advantage in solving self-consistently  Eq.~(\ref{eq:ABavg}) in comparison to solving Eq.~(\ref{eq:dFmf}) as it requires no evaluation of numerical derivatives, which can introduce errors of order of grid separation ($\sim 1/N$) and allows for finding completely unrestricted solutions \cite{thermal3}. We take $J_1=1$ as our unit of energy and, the distance between $u$ to next $u$ sub-lattices, as our unit of length.

The minimization technique we use is as follows. First we choose a set of mean-field parameters $\mathcal{O}$ depending on the possible ground-state, which needs to be taken carefully for convergence. In the initial step, for this set of $\mathcal{O}$s, we scan for allowed values of the chemical potentials $\mu_u,\mu_v$, such that the constraint, Eq.~(\ref{eq:mu}), is satisfied on both sub-lattices, in the ground-state. In the next step, we evaluate the modified mean-field parameters using Eq.~(\ref{eq:ABavg}) (which is simplified by using Eq.~(\ref{eq:gammags})), and again we find appropriate $\mu_u,\mu_v$, such that the constraint, Eq.~(\ref{eq:mu}), is satisfied on both sub-lattices, in the modified ground-state. This procedure continues until the mean-field parameters as well as well the chemical potentials converge up to a value of tolerance.  We first obtain the solutions for $J_{\chi}=0$, and then we use these solutions as initial seeds for solutions with small $J_{\chi}$, and follow the same procedure with successively increasing $J_{\chi}$. In our case, the tolerance on mean-field parameters at least $\sim$ $10^{-6}$. 

We distinguish different phases of the ground-state by following properties. First, we call a state gapless, if the gap in the spectrum is less-than $1/N$, as there is always a finite-size gap present in our system even though the state can be gapless in the thermodynamic limit. Next, we look for the symmetries of the converged mean-field parameters, which can predict the nature of the ground-state based on projective symmetry ground analysis, which we present later. In the gapless state, the momentum where the spectrum is minimum dictates the ordering vector and thus the nature of the long range order. For the gapped state, we also compute the Chern number of one of the excitations bands to distinguish between a trivial QSL state (we call it GSL) or VBC, where the Chern number is zero, from a CZSL state (with non-zero Chern number). Finally, we also compute the static spin-spin correlation in the ground-state. How fast this correlation decays as a function of the distance between two sites can differentiate the nature of the ground-state.

\begin{figure*}[t]
	\centering
	\includegraphics[width=1.0\linewidth]{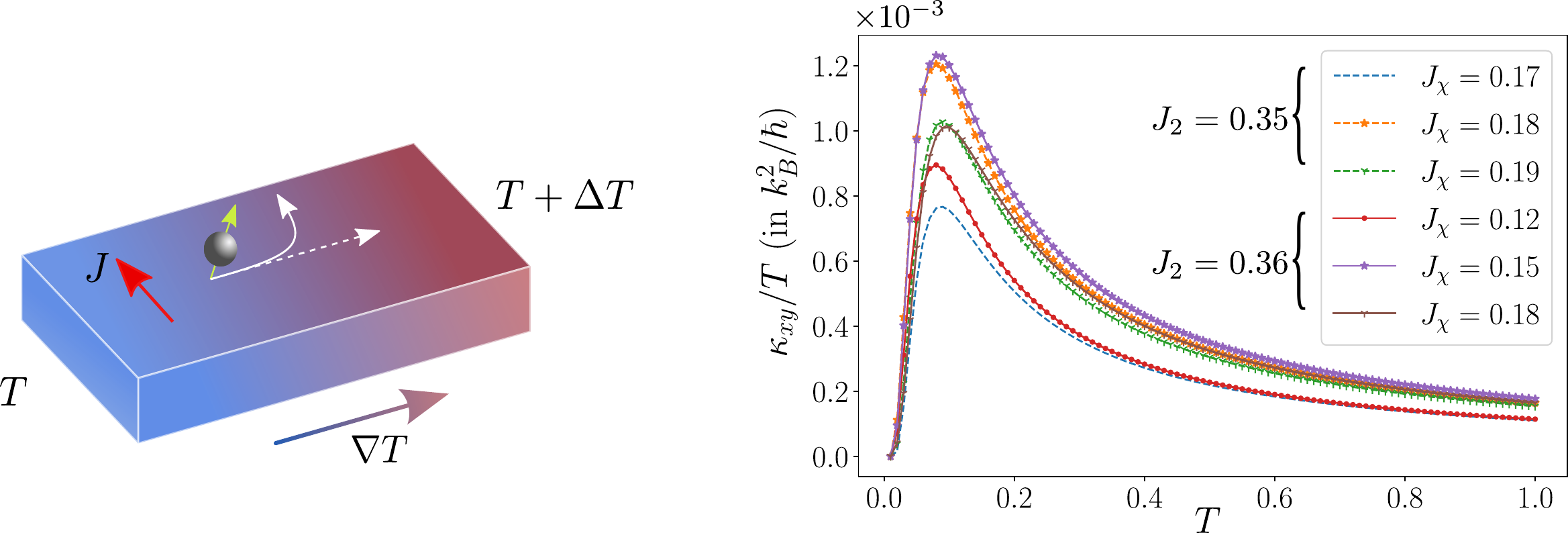}
	\caption{left: schematic picture of the set up where thermal Hall current carried by spinons in the presence of a longitudinal magnetic field, where the role of the magnetic field is played by the scalar chiral coupling in our system. Right: the thermal Hall coefficient ($\kappa_{xy}$) as a function of the temperature for the states with $C\neq0$. A critical $T$ for the non-zero $\kappa_{xy}$ reflects the fact that the spin bands are gapped. In the states with $C=0$, the Hall coefficient is zero, in comparison.}
	\label{fig:thermalhall2}
\end{figure*}

\subsubsection*{Projective Symmetry of Ansatz}
The original idea of projective symmetry groups (PSG) classification for spin-liquids was introduced by Wen and collaborator~\cite{wen2002quantum,wen,zhou2002quantum}, in the context of Schwinger-Fermion approach. PSG analysis in Schwinger-Boson approach was extended by Wang et all~\cite{subir4}. Study of PSG provides the allowed symmetries and sign structures of the mean field Ansatz. In disordered phase we want our mean-field state to obey the underlying microscopic symmetries of the spin model. For honeycomb lattice this symmetry transformations are lattice translations, point group symmetries (ie $C_{3}$ rotation and  reflections), spin rotation symmetry  and time reversal symmetry. Additionally, for the case of Schwinger-bosons,  under the local U(1)  transformation 
\begin{equation}\label{eq:localg}
	b_{\vec{r}\sigma} \rightarrow e^{i\phi(\vec{r})}b_{\vec{r}\sigma},
\end{equation}
under which the mean-field ansatz transform as:
\begin{eqnarray}
	A_{ij} \rightarrow e^{-i\phi(i)-i\phi(j)}A_{ij} \ \ B_{ij} \rightarrow e^{+i\phi(i)-i\phi(j)}B_{ij}.
\end{eqnarray}
all the physical observables should remain invariant. But a subset of this U(1) transformation keeps the ansatz themselves invariant. The set of all transformations that keep the ansatz invariant form the PSG. The set of the elements of PSG that are of kind Eq.~(\ref{eq:localg}), 
form a group called Invariant Gauge Group (IGG)~\cite{wen2002quantum}. For Honeycomb lattice with both nonzero $A_{ij}$ and $B_{ij}$, this IGG is simply $Z_{2}$~\cite{subir5}. For the honeycomb lattice, such PSG classification found to give two distinct spin liquid states classified as 0 and $\pi$ flux states~\cite{subir5}. The spin liquid states we found from numerical simulations, starting from completely unrestricted ansatz, matches with the 0 flux states mentioned above.   
 
The symmetries of the ansatz that identifies the nature of the ground-state are following~\cite{subir5,thermal3}
\begin{itemize}
\item In Spin Liquid State and Neel state:
\begin{align}\label{eq:gslmf}
	&A_{11}=A_{12}=A_{13}=A \nonumber \\
	&A^{u}_{21}=A^{u}_{22}=A^{u}_{23}=A^{v}_{21}=A^{v}_{22}=A^{v}_{23}=0  \nonumber\\
	&B_{11}=B_{12}=B_{13}=0  \nonumber\\
	&B^{u}_{21}=B^{u}_{22}=B^{u}_{23}=B^{v}_{21}=B^{v}_{22}=B^{v}_{23}=B.
\end{align}
\item In VBC State:
\begin{align}
&A_{11} \neq A_{12} = A_{13}, ~A^{u}_{21} \neq A^{u}_{22} =- A^{u}_{23}\nonumber\\ 
&A^{v}_{21} \neq A^{v}_{22} =- A^{v}_{23} \nonumber\\
&B_{11} \neq B_{12} = B_{13}, ~B^{u}_{21} \neq B^{u}_{22} = B^{u}_{23} \nonumber\\ 
&B^{v}_{21} \neq B^{v}_{22} = B^{v}_{23}.
\end{align}
\item In Spiral State:
\begin{align}
&	A_{11}=A_{12} \neq A_{13}, ~A^{u}_{21} \neq A^{u}_{22} =- A^{u}_{23} \nonumber\\
&A^{v}_{21} \neq A^{v}_{22} =- A^{v}_{23} \nonumber\\
&	B_{11} = B_{12} \neq B_{13},~	B^{u}_{21} \neq B^{u}_{22} = B^{u}_{23} \nonumber\\
&B^{v}_{21} \neq B^{v}_{22} = B^{v}_{23}.
\end{align}
\end{itemize}
The broken time-reversal symmetry of state give rise to non-vanishing imaginary part of the mean-field parameters we obtain~\cite{subir4,chiralpsg}, in the chiral state. As the $A$ and $B$, in Eq.~(\ref{eq:gslmf}), are both non-zero in numerical finding, we identify the GSL state as a $Z_2$ quantum spin-liquid~\cite{subir4}.

\subsubsection*{$C_3$-symmetry breaking order parameter}
In the intermediate spin disordered region, in the VBC state, the spin rotational symmetry $SU(2)$ and transnational symmetries are intact, but it may break the $C_{3}$ rotational symmetry of the lattice. Following Okumura \textit{et al.},~\cite{vbc} we define a $C_{3}$ rotational symmetry breaking order parameter,
\begin{align}
&\psi_{3}=p_{1}\vec{a}_{1}+p_{2}\vec{a}_{2}+p_{3}\vec{a}_{3},\label{eq:C3}\\
&{\rm with,~}p_{i}=J_{1}(B_{1i}^{2}-A_{1i}^{2}).\nonumber
\end{align} 
$p_{\alpha} \ (\alpha=1,2,3)$ are nothing but the bond energies corresponding to nearest-neighbor bonds $\vec{a}_{\alpha} \ (\alpha=1,2,3)$. This order parameter is zero as long as the bond energies remain same along three different direction.

\section{Numerical Results}\label{sec:numres}
Without application of the scalar chiral term (i.e, $J_{\chi}=0$), numerically we find, for $J_{2} \leq 0.22$, the ground-state is gapless (defined as a gap less than 1/$N$), with N\'{e}el order and spiral magnetic order is also found for larger value of $J_{2} >0.4$.  A gapped phase is found in the intermediate range $0.22<J_{2}<0.4$ between the N\'{e}el and the spiral order. Within a range of $0.37 \leq J_{2}<0.4$ we find the staggered valence bond crystal (VBC) phase, with non-zero $C_3$ symmetry breaking order parameter, Eq.~(\ref{eq:C3}). We call the rest of the gapped region GSL state. These findings match with previous studies~\cite{neel,thermal3}.

In Fig.~\ref{fig:dispersion}, we show the lower spinon (particle) bands, without the application of $J_{\chi}$ in all the four different phases. The Brillouin-zone is from $-\pi$ to $\pi$ in both momentum, measured along the directions along the reciprocal translation vectors. The dispersion in the N\'{e}el ordered phase shows the characteristic minima (with gap $<1/N$) at the $(0,0)$ momentum, which shifts away from this point in the case of the spiral ordered state. The spectrum for the GSL and the VBC states are gapped with different positions of minima in the band. The gap in the spectrum for different phase is shown in the Fig.~\ref{fig:gap1}, where we also show, in the inset of the same figure, the sudden rise in the $\psi_3$ order-parameter in the VBC state.

Within the GSL state, as the the $J_{\chi}$ is increased, we find, beyond a certain value of $J_{\chi}$, either the state becomes gapless with N\'{e}el ordering, or the bands acquire non-zero Chern number, which we identify as a CZSL state. It is important to note that, in CZSL state, spinon bands remain gapped but with increasing perturbation ($J_{\chi}$), the particle bands themselves come closer leading to topological phase transition for a critical $J_{\chi}$. If we start instead from a VBC state, for a critical perturbation we also observe a topological transition in the spinon bands (the ground-state still remains gapped). Interestingly, we observe that, if we start in GSL state, we end up with a Chern number $C=1$, whereas, if we start from a VBC state, we obtain a Chern number $C=2$, after the topological transition.

In the Fig.~\ref{fig:phase} (a), we show the full phase-diagram including the $J_{\chi}$ perturbation which leads to possible CZSL state, characterized by non-zero Chern number. We also show the static spin-spin correlation, defined in the Sec.~\ref{sec:ssf} in Fig.~\ref{fig:phase} (b) and (c) for the phases without $J_{\chi}$ and with $J_{\chi}$, respectively. From these logarithmic-scaled plots, it is evident that the spin-spin correlation, $|\langle \vec{\hat{S}}_{0}  \cdot \vec{\hat{S}}_{R} \rangle |$, decays at a much faster rate, as a function of the distance $R$, in the GSL, CZSL and VBC state in comparison to magnetically ordered states, which is expected.

With increasing $J_{\chi}$, at a critical $J_{\chi}$, there is a topological transition to a non-zero Chern number ($C$) state, which can also be seen from the Berry curvatures of the spinon bands. When $C\neq0$, the symmetry the Berry curvature is lost, i.e, $\Omega(\vec{q})\neq\Omega(-\vec{q})$. The plot of Berry curvature is shown in Fig.~\ref{fig:berry1}, before and after such a topological transition.

Finally, in Fig.~\ref{fig:thermalhall2}, we show the thermal Hall coefficients in the states with $C\neq0$, which peaks to an appreciable value at a temperature equal to the gap in the lower spinon band. Due to the preserved symmetry of the Berry curvature, the Hall coefficients is vanishingly small in the case of the state with $C=0$. 

\section{Discussion}\label{sec:sum}
In conclusion, we investigated within the Schwinger-Boson mean-field theory, the phase transition from a gapped $Z_{2}$ quantum spin-liquid, in a $J_1$-$J_2$ Heisenberg spin-1/2 system in a honeycomb lattice, to a chiral $Z_{2}$ spin liquid (CZSL) phase under the presence of time-reversal symmetry breaking scalar chiral interaction (with amplitude $J_{\chi}$). The CZSL state is characterized by time-reversal broken mean-field parameters, non-trivial Chern bands for excitations and lack of long-range magnetic order. In this CZSL phase we find non-trivial Chern number of the spinon bands leads to large thermal Hall coefficient. 

The study is limited by the finite-size effects in the spectrum, and a comparison to larger system size is left for future study. Similarly, the topological invariant computation can be erroneous in situations when the gap between the spinon bands is too small. The detailed nature of the $C\neq0$ is also left unexplored where one expects topological transitions with further increase of the value of $J_{\chi}$.

Among possible materials for QSL states in Mott insulating states in honeycomb lattice includes inorganic materials such as \ce{Na_{2}Co_{2}TeO_{6}} \cite{ralko1}, \ce{BaM_{2}(XO_{4})_{2}} (with \ce{X=As}) \cite{ralko2}, \ce{Bi_{3}Mn_{4}O_{12}(NO_{3})} \cite{ralko3} and \ce{In_{3}Cu_{2}VO_{9}} \cite{ralko4}, where the magnitudes of spin varies from $S=1/2$ in \ce{BaM_{2}(XO_{4})_{2}} for \ce{M=Co} to $S=1$ for \ce{M=Ni} (with \ce{X=As}) and $S=3/2$ in \ce{Bi_{3}Mn_{4}O_{12}(NO_{3})}. Spin-orbit coupled materials, such as, $\ce{In_{3}Cu_{2}VO_{9}}$ has also been recently explored~\cite{ralko4} where the Cu ions form a honeycomb structure with spin-1/2 local moments, which is also a possible candidate material. Another possible avenue of realizing spin systems are cold-atomic experiments where strongly correlated systems have been explored in recent times~\cite{jepsen2020spin,goldman2016topological,aidelsburger2015measuring,ebadi2021quantum}.

\section{Acknowledgments}
R.M. thanks useful communication with Shubhayu Chatterjee (UC Berkeley), Arnaud Ralko (N\'{e}el inst., CNRS Grenoble) and Jaime Merino (UAM, Madrid). A.K  acknowledges support from the SERB (Govt. of India) via saction no. ECR/2018/001443, DAE (Govt. of India ) via sanction no. 58/20/15/2019-BRNS, as well as MHRD (Govt. of India) via sanction no. SPARC/2018-2019/P538/SL. R.M. acknowledges the CSIR (Govt. of India) for financial support. R.K. acknowledges funding under PMRF scheme (Govt. of India). We also acknowledge the use of HPC facility at IIT Kanpur.

\bibliography{reference.bib}
\beginsupplement
\appendix{}
\subsection*{Appendix A: Berry curvature and U(1)-link variable}\label{app:U1}
Here we briefly summarize the method of Berry-curvature computation, especially for the Bosonic case, following Ref.~\onlinecite{berry6} and Ref.~\onlinecite{berry7}.  We first consider a  two dimensional \textit{fermionic} system with the Brillouin zone defined by $0 \leq q_{\mu} < 2\pi/a_{\mu}$ ($\mu=1,2$ with some integer $a_{\mu}$). As the Hamiltonian $H(q_{1},q_{2})$ is periodic in both directions, $H(q_{1},q_{2})=H(q_{1}+2 \pi/a_{1},q_{2})=H(q_{1},q_{2}+ 2 \pi/a_{2})$.

The Berry connection $A_{\mu}(q) \ (\mu=1,2)$ and the corresponding field-strength $F_{12}(q)$, for the $n^{\rm th}$ band, are given by
\begin{equation}
	A^n_{\mu}(q)=\bra{n(q)} \partial_{\mu} \ket{n(q)},
\end{equation}
\begin{equation}
	F^n_{12}(k)= \partial_{1} A^n_{2}(q) -\partial_{2} A^n_{1}(q),
\end{equation}
where $\ket{n(q)}$ being a normalized wave function of the $n^{\rm th}$ Bloch band such that,
\begin{equation}
	H(q)\ket{n(q)}=E_{n}(q)\ket{n(q)}.
\end{equation}
In the expression above the derivative $\partial_{\mu}$ stands for $\partial_{q_{\mu}}$. We assume that there is no degeneracy for the $n^{\rm th}$ state.

The Berry curvature is computed as following. First we discretize the  Brillouin zone as following:
\begin{equation}
	q_{l}=(q_{j_{1}},q_{j_{2}}), \ \ q_{j_{\mu}}=\dfrac{2\pi j_{\mu}}{a_{\mu}N_{\mu}}, \ \ (j_{\mu}=0,....,N_{\mu}-1),
\end{equation}
with discretization $\delta q_{\mu} = 2\pi/a_{\mu}N_{\mu}$. It is also assumed that the state $\ket{n(\vec{q})}$ is periodic on the lattice,
\begin{equation}
	\ket{n(\vec{q}+N_{\mu}\hat{\mu})}=\ket{n(k_{l})},
\end{equation}
where $\hat{\mu}$ is a vector in the direction $\mu$ with magnitude $2\pi/(a_{\mu} N_{\mu})$. We define the $U_{1}$ link variable for the $n^{\rm th}$ band as,
\begin{equation}
	U^n_{\hat{\mu}}(\vec{q}) \equiv \langle n(\vec{q})| n(\vec{q}+\hat{\mu})\rangle/N^n_{\hat{\mu}}(\vec{q})
\end{equation}
where,
\begin{equation}
	N^n_{\hat{\mu}}(\vec{q}) \equiv |\langle n(\vec{q})|
	n(\vec{q}+\hat{\mu})\rangle|.
\end{equation}
The link variables are well defined as long as $N^n_{\hat{\mu}}(q_{l}) \neq 0$, which can always be assumed to be the case (one can avoid a singular point by infinitesimal shift of the lattice). The field-strength is then numerically approximated by
\begin{equation}{\label{eqn54}}
F^n_{12}(\vec{q})\delta q_1 \delta q_2 \approx \log_{e} U^n_{1}(\vec{q})U^n_{2}(\vec{q}+\hat{1})U^n_{1}(\vec{q}+\hat{2})^{-1}U^n_{2}(\vec{q})^{-1}.
\end{equation}
with,
\begin{equation}
	-\pi < \dfrac{1}{i} F^n_{12}(\vec{q}) \delta q_1 \delta q_2  \leq \pi.
\end{equation}
Field-strength is defined within the principle branch of the logarithm specified in Eq.~(\ref{eqn54}). It should also be noted that field strength is gauge-invariant. The Berry curvature is expressed in terms of the field-strength as
\begin{align}
	\Omega^n(\vec{q}) = -i F_{12}(\vec{q}).
\end{align}
Finally the Chern number on the lattice corresponding to the $n^{\rm th}$ band is defined as,
\begin{equation}
C_{n} \equiv \dfrac{1}{2\pi i}\sum _{\vec{q}}F_{12}(\vec{q}) \delta q_1 \delta q_2 .
\end{equation}

\subsubsection*{For Bosonic case}
To accommodate the commutation relations among the bosonic operators, the generalized eigenvalue equation in case of a bosonic Hamiltonian $M$  is written as,
\begin{equation}
	M(q)\ket{n(q)}=E(q)\rho_{3} \ket{n(q)}, 
\end{equation}
as a consequence the inner product in U(1)-link variable has the form \cite{berry7},
\begin{equation}
	U_{\hat{\mu}}(\vec{q}) \equiv \bra{n(\vec{q})} \rho_{3} \ket{n(\vec{q}+\hat{\mu})}/N_{\hat{\mu}}(\vec{q})
\end{equation}
where,
\begin{equation}
	\emph{N}_{\hat{\mu}}(\vec{q}) \equiv |\bra{n(\vec{q})}\rho_{3}\ket{n(\vec{q}+\hat{\mu})}|.
\end{equation}
$E(q)$ has eigenvalues of the form,
 \begin{equation}
	(\epsilon_{\vec{q},\uparrow}, \epsilon_{\vec{q},\downarrow}, -\epsilon_{\vec{q},\uparrow}, -\epsilon_{\vec{q},\downarrow})
\end{equation}
For particle/hole bands the eigenvector $\ket{n(\vec{q})}$ is normalized as follows,
\begin{align}
&	\bra{n^{\rm particle}(\vec{q})} \rho_{3} {\ket{n^{\rm particle}(\vec{q})}}=1,\\
&	\bra{n^{\rm hole}(\vec{q})} \rho_{3} {\ket{n^{\rm hole}(\vec{q})}}=-1.
\end{align}

\end{document}